\documentclass[10pt,a4paper]{article}

\usepackage[utf8]{inputenc}
\usepackage[english]{babel}
\usepackage{amsmath}
\usepackage{amsfonts}
\usepackage{amssymb}
\usepackage{makeidx}
\usepackage{graphicx}
\usepackage{arydshln}
\usepackage{cancel}
\usepackage{tcolorbox}
\usepackage[colorlinks=true, linkcolor=blue, citecolor=cyan, urlcolor=blue]{hyperref}
\usepackage{dcolumn}
\usepackage{bm}
\usepackage{colortbl}
\usepackage{authblk}

\usepackage{array}
\usepackage{makecell}
\usepackage{breqn}

\makeatletter
\preto\maketitle{%
  \begingroup\lccode`~=`,
  \lowercase{\endgroup
  \let\saved@breqn@active@comma~
  \let~}\active@comma 
}
\appto\maketitle{%
  \begingroup\lccode`~=`,
  \lowercase{\endgroup
  \let~}\saved@breqn@active@comma 
}
\makeatother

\newtheorem{proposition}{Proposition}
\newtheorem{theorem}{Theorem}
\newtheorem{definition}{Definition}

\def\be{\begin{equation}}
\def\ee{\end{equation}}
\def\X{\bm{X}}

\newcommand{\ie}{\textit{i.e. }}
\newcommand{\eg}{\textit{e.g. }}

\begin{document}


\title{On the equivalence between the Kinetic Ising Model and discrete autoregressive processes}

\author[1,2,3]{Carlo Campajola}
\author[1,4]{Fabrizio Lillo}
\author[1]{Piero Mazzarisi}
\author[4]{Daniele Tantari}
\affil[1]{Scuola Normale Superiore, piazza dei Cavalieri 7, 56126 Pisa, Italy}
\affil[2]{UZH Blockchain Center, University of Z\"{u}rich, R\"{a}mistrasse  71,  8006  Z\"{u}rich, Switzerland}
\affil[3]{URPP Social Networks, University of Z\"{u}rich, Andreasstrasse 15, 8050 Z\"{u}rich, Switzerland}

\affil[4]{Department of Mathematics, University of Bologna, piazza di Porta San Donato 5, 40126 Bologna, Italy}



\date{\today}
\maketitle

\begin{abstract}
Binary random variables are the building blocks used to describe a large variety of systems, from magnetic spins to financial time series and neuron activity. In Statistical Physics the Kinetic Ising Model has been introduced to describe the dynamics of the magnetic moments of a spin lattice, while in time series analysis discrete autoregressive processes have been designed to capture the multivariate dependence structure across binary time series. In this article we provide a rigorous proof of the equivalence between the two models in the range of a unique and invertible map unambiguously linking one model parameters set to the other. Our result finds further justification acknowledging that both models provide maximum entropy distributions of binary time series with given means, auto-correlations, and lagged cross-correlations of order one. We further show that the equivalence between the two models permits to exploit the inference methods originally developed for one model in the inference of the other.
\end{abstract}



\section{Introduction}

The dynamics of a large variety of systems, from Physics to Economics and Finance, can be represented as time series of binary variables. The most telling examples are the {\it spin systems} in Statistical Physics, where magnetic moments of the particles in a lattice are described as two-state variables, or {\it binary time series} in Quantitative Finance, capturing for instance the occurrence of extreme events of prices \cite{calcagnile2018collective,hong2009granger}, or buy and sell orders in the order book of financial markets \cite{bouchaud2002statistical}. Different models have been introduced to capture the multivariate interaction structure of such binary systems, in particular the {\it Kinetic Ising Model} (KIM)  \cite{fredrickson1984kinetic} in Physics and the {\it discrete autoregressive processes} \cite{jacobs1978discrete,jacobs1983stationary} in time series analysis, together with the (Markovian) multivariate generalization recently introduced by \cite{mazzarisi2020tail}, namely the{\it Vector Discrete AutoRegressive Process} VDAR(1). 
In this paper we prove analytically that, under some condition, the KIM is equivalent to the VDAR(1) model.
Furthermore it is well known that the Ising model, in both static \cite{schneidman2006weak} and kinetic \cite{marre2009prediction} version, is a maximum entropy model, given mean magnetizations and pairwise correlations (at lag one in the kinetic case), see also \cite{jaynes1982rationale,presse2013principles,marcaccioli2020correspondence}, and, among other aspects, maximum entropy arguments can be used to define without ambiguity the temperature in such nonequilibrium spin systems \cite{sastre2003nominal}.
Here, by exploiting the equivalence between the two models, we prove also that the Markov chain associated with the vector discrete autoregressive process VDAR(1) can be interpreted as the maximum entropy distribution of binary random variables  with given {\it means}, {\it auto-correlations}, and {\it lagged cross-correlations} (of order one).   Thus, the KIM and the VDAR(1) should be preferable to other models in absence of prior information on other metrics, following the principle of maximum entropy.

The Kinetic Ising Model \cite{fredrickson1984kinetic, derrida1987exactly, crisanti1988dynamics, sides1998kinetic}, was originally proposed as the out of equilibrium version of the classical Ising spin glass \cite{ising1925beitrag, edwards1975theory, kirkpatrick1978infinite} to describe a Markovian dynamics of {\it spins} $\sigma_t^i$, \ie binary random variables taking values $-1$ and $1$, interacting with each other according to some generic matrix of couplings. The KIM has found countless applications in many contexts such as neuroscience \cite{roudi2011mean, capone2015inferring}, computational biology \cite{imparato2007ising}, machine learning \cite{coolen1993dynamics, dunn2013learning, campajola2019inference}, and economics and finance \cite{bouchaud2013, campajola2020unveiling, campajola2020modelling}.

In mathematical terms, the KIM is a logistic regression model 
specified by the following transition probability for a set of $N$ spins $\boldsymbol{\sigma}_t\equiv\{\sigma_t^i\}_{i=1,...,N}$, 

\begin{dmath}\label{eq:transprob}
    p_{KIM}(\boldsymbol{\sigma}_{t} \vert \boldsymbol{\sigma}_{t-1}, \bm{J}, \bm{h}) =  Z^{-1}_{t-1} \exp \left( \sum_{ i, j =1}^N \sigma^i_{t} J_{ij} \sigma^j_{t-1} + \sum_{i=1}^N \sigma^i_{t} h_i \right)
\end{dmath}

where $\bm{J}\equiv\{J_{ij}\}_{i,j=1,...,N}$ is a matrix of real-valued parameters giving the multivariate auto-regressive structure of the model or, equivalently, representing the couplings between spins, $\bm{h}\equiv\{h_i\}_{i=1,...,N}$ is a set of variable-specific parameters representing the external magnetic fields associated with each spin, and $Z_{t-1}$ is the partition function $Z_{t-1} = \prod_{i=1}^N 2 \cosh (\sum_j J_{ij} \sigma^j_{t-1} + h_i)$ guaranteeing that the probability distribution is properly normalized.

The KIM of Eq. (\ref{eq:transprob}) is a {\it maximum entropy} \cite{jaynes1957information, barnett2013information} model for a set of binary random variables which display on average given {\it means}, and both {\it auto-} and (lagged) {\it cross-correlations}. For the sake of clarity and in preparation to the section below, let us move from the spin variables $\sigma_t^i\in\{-1,1\}$ to the binary variables $X_t^i\in\{0,1\}$.

Given a set of $N$ binary variables $\X_t\equiv\{X_t^i\}_{i=1,...,N}^{t=1,...,T}$, let us consider the following metrics,
\be\label{eq:mean}
2\sum_tX_t^i,\:\:\forall i=1,...,N,
\ee
\be\label{eq:corr}
2\sum_tX_t^iX_{t-1}^j+(1-X_t^i)(1-X_{t-1}^j),\:\:\forall i,j=1,...,N,
\ee
related (under stationarity conditions) to the mean of the binary random variables and the correlation between them, respectively. 

The metric (\ref{eq:corr}) for $i=j$ is known as {\it stability} \cite{hanneke2010discrete}, which is connected with the sample auto-correlation of a binary sequence, \ie $\sum_{t}X_t^iX_{t-1}^i$. Similarly, when $i \neq j$ the metric is related to lagged cross-correlations. The maximum entropy probability distribution of $\X_1,\X_2,...,\X_T$, \ie the one maximizing the entropy $-\sum_{\X_1,...,\X_T}p(\X_1,...,\X_T)\log p(\X_1,...,\X_T)$ while preserving on average some given values for the metrics (\ref{eq:mean}) and (\ref{eq:corr}), has transition probability (by assuming a given initial condition $\X_0$ and exploiting the Markov property)  

\begin{equation}\label{eq:tpexp}
p(\boldsymbol{X}_t \vert \boldsymbol{X}_{t-1};  \bm{J}, \bm{h}) = \prod_{i=1}^N \frac{\exp\left[2X_t^i \left( h_i + \sum_{j=1}^N J_{ij} (2X_{t-1}^j - 1)\right)\right]}{1+\exp\left[2\left(h_i+\sum_{j=1}^N J_{ij}(2X_{t-1}^j - 1)\right)\right]}
\end{equation}

where $\bm{J}=\{J_{ij}\}_{i,j=1,...,N}$ and $\bm{h}=\{h_i\}_{i=1,...,N}$ are $N^2+N$ Lagrange multipliers solving the maximum entropy problem \cite{jaynes1957information, park2004statistical}. It is trivial to show that the transition probability of the Kinetic Ising Model (\ref{eq:transprob}) can be stated equivalently as the maximum entropy probability (\ref{eq:tpexp}) in terms of the binary variables $\boldsymbol{X}_t$ $\forall i,t$, through the relation $X_t^i = \frac{1 + \sigma^i_t}{2}$ (with the same parameters $\bm{J}$ and $\bm{h}$).

The VDAR(1) model describes the dependence structure of a set of binary random variables which has Markov property and Bernoulli marginal distribution. It has been proposed originally in its univariate version \cite{jacobs1978discrete} and followed by several extensions such as the Discrete AutoRegressive Moving Average (DARMA) model \cite{jacobs1983stationary} and recently proposed in its multivariate formulation \cite{mazzarisi2020tail}, the VDAR model. Models from this family have seen applications in genetics \cite{dehnert2003discrete}, queueing theory \cite{kim2008mean}, temporal networks \cite{williams2019effects} and recently in financial systems, as methods to forecast order flows \cite{taranto2014adaptive} or to identify preferential lending between banks \cite{mazzarisi2020dynamic}.

In terms of the $N$ binary variables $\X_t\equiv\{X_t^i\}_{i=1,...,N}$ with $X_t^i\in\{0,1\}$ (and initial condition $\X_0$), the VDAR(1) process describes the evolution of $X_t^i$ as
\be\label{eq:vdar}
X_t^i=V_t^iX_{t-1}^{A_t^i}+(1-V_t^i)Z_t^i
\ee
with $V_t^i\sim\mathcal{B}(\nu_i)$ a Bernoulli random variable with parameter $\nu_i\in[0,1]$, $A_t^i\sim\mathcal{M}(\lambda_{i1},...\lambda_{iN})$ a multinomial random variable taking integer value in \{1,...,N\}, with parameters $\lambda_{i1},...\lambda_{iN}$ such that $\sum_{j=1}^N\lambda_{ij}=1$, and $Z_t^i\sim\mathcal{B}(\chi_i)$ with $\chi_i\in[0,1]$. In other words, the VDAR(1) process captures the (multivariate) mechanism of copying from the past: with probability $\nu_i$, $X_t^i$ is copied from the past and, in this case, $\lambda_{ij}$ is the probability that $X_t^i$ is equal to $X_{t-1}^j$ (including also the past itself with probability $\lambda_{ii}$); otherwise, with probability $1-\nu_i$, $X_t^i$ is not copied and is instead sampled according to a Bernoulli marginal with probability $\chi_i$. Hence, the VDAR(1) model describes $N$ binary random variables with both Markov property and some autoregressive dependency structure, similarly to the KIM. The model is formalized by the transition probability
\begin{dmath}\label{eq:tpvdar1}
p_{VDAR}(\boldsymbol{X}_{t} \vert \boldsymbol{X}_{t-1}; \boldsymbol{\pi}) = \prod_{i=1}^N \left[ \nu_i \left( \sum_{j=1}^N \lambda_{ij} \delta_{X_t^i,X_{t-1}^j}\right) +(1-\nu_i)(\chi_i)^{X_t^i}(1-\chi_i)^{1-X_t^i}\right]
\end{dmath}
where $\delta_{X_t^i,X_{t-1}^j}$ is the Kronecker delta, $\boldsymbol{\pi} = \{\lbrace \nu_i \rbrace, \lbrace \lambda_{ij}\rbrace, \lbrace \chi_i \rbrace\}_{i,j=1,...,N}$. 

Notice that the model has $N^2+N$ parameters, exactly as the Kinetic Ising Model.
It is thus immediate to ask the question whether a mapping between the two models exists, as well as finding under which conditions the two models can be considered equivalent. In the following, we indicate the KIM model as $\lbrace \lbrace \boldsymbol{X}_t \rbrace, p_{KIM}, \boldsymbol{\theta} \rbrace$ with set of parameters $\boldsymbol{\theta} \equiv (\bm{J},\bm{h})$, while the VDAR(1) model is summarized as $\lbrace \lbrace \boldsymbol{X}_t \rbrace, p_{VDAR}, \boldsymbol{\pi} \rbrace$. Calling $\Theta = \mathbb{R}^{N \times N} \times \mathbb{R}^N$ the space of all possible KIM parameters $\boldsymbol{\theta}$ and $\Pi$ the space of all possible VDAR(1) parameters $\boldsymbol{\pi}$,
\begin{definition}
The KIM and the VDAR(1) models are said to be  equivalent on $(\hat{\Theta},\hat{\Pi})$ if there exist an unique invertible map $f:\hat{\Pi}\subseteq\Pi \rightarrow \hat{\Theta}\subseteq\Theta$ such that  
$$p_{KIM}(\boldsymbol{X}_t \vert \boldsymbol{X}_{t-1}; f(\boldsymbol{\pi})) = p_{VDAR}(\boldsymbol{X}_t \vert \boldsymbol{X}_{t-1}; \boldsymbol{\pi})$$
for any $\boldsymbol{X}_t$ and $\boldsymbol{X}_{t-1}$.
\end{definition}

\section{Model equivalence}

Before stating the main theorem, let us show that this mapping exists in the trivial cases of $N=1$ and $N=2$. For $N=1$, both $J$ and $h$ are scalar parameters, while the DAR(1) model, namely the univariate version of VDAR(1), has two parameters, \ie $\nu$ and $\chi$ ($\lambda = 1$ by design, since we can copy only the past value of the single variable), thus it is trivial to prove that

\begin{subequations}
\begin{equation}\label{thetaeq}
h = \frac{1}{4} \log \left( \frac{\frac{\chi}{1-\chi}+\nu}{\frac{1}{\chi}-(1-\nu)}\right)
\end{equation}
\begin{equation}\label{mueq}
J = \frac{1}{4}\log\left(1+\frac{\nu}{(1-\nu)^2\chi(1-\chi)}\right)
\end{equation}

\end{subequations}

One can notice that here $J$ is strictly positive as long as $\nu, \chi > 0$, while it is $J=0$ if and only if $\nu = 0$: this suggests that the VDAR(1) model is indeed a restricted version of the KIM, with the elements of the coupling matrix restricted to positive values. Intuitively, in the KIM  $J_{ij}<0$ implies that spin $i$ tends to take the opposite value of the past state of $j$, whereas the VDAR model describes only positive (or zero) correlations.

When $N=2$, both models have $6$ free parameters, three parameters associated with each variable (spin) $i=1,2$, thus one can map the two models by considering three independent configurations of $\boldsymbol{X}_{t-1}$, \eg $\{X_{t-1}^1,X_{t-1}^2\}=\{1,1\},\{0,1\},\{0,0\}$ and one possible realization of $X_t^i$, \eg $X_t^i=1$, for both cases $i=1$ and $i=2$. Then, by matching the transition probabilities for the two models, one obtains the following system

\begin{dmath}\label{dar2problem}
\begin{pmatrix} 
1 & 1 & -1 \\
1 & -1 & -1 \\
-1 & -1 & -1 \\
\end{pmatrix}
\begin{pmatrix} 
J_{i1} \\
J_{i2} \\
h_i \\
\end{pmatrix}
 =\frac{1}{2}
\begin{pmatrix} 
\log \left(\frac{1}{(1-\nu_i)\chi_i}-1\right) \\
\log \left(\frac{1}{\nu_i(1-\lambda_{i})+(1-\nu_i)\chi_i}-1\right) \\
\log \left(\frac{1}{\nu_i+(1-\nu_i)\chi_i}-1\right) \\
\end{pmatrix}
\end{dmath}
\noindent $\forall$ $i = 1,2$.

Hence, there exists a {\it unique} mapping $f:\Pi \rightarrow \Theta$ as long as the linear system of equations (\ref{dar2problem}) admits a solution in the domain of parameters: (i) the solution exists because the matrix in the left hand side of (\ref{dar2problem}) is invertible, then (ii) the mapping $f$ admits the  inverse $f^{-1}:\Theta\vert_{J_{ij}\geq 0} \rightarrow \Pi$ in the restricted codomain $J_{ij}\geq 0$ $\forall i,j$ (as can be verified by simple computations).

Given these premises, we can now move to the main result of this paper, by stating

\begin{theorem}\label{th:equiv}
The Kinetic Ising Model $\lbrace \lbrace \boldsymbol{X}_t \rbrace, p_{KIM}, \boldsymbol{\theta} \rbrace$ is equivalent to the VDAR(1) model $\lbrace \lbrace \boldsymbol{X}_t \rbrace, p_{VDAR}, \boldsymbol{\pi} \rbrace$ if and only if $J_{ij} \geq 0$ $\forall i,j$.
\end{theorem}

In order to prove the theorem above, let us first prove the existence of a map $f:\Pi \rightarrow \Theta$ from the VDAR(1) model to the KIM, for any set of parameters $\bm{\pi}\in\Pi$. In particular, this map is {\it unique} because of the linearity of the mapping problem, see below. Second, we prove that the mapping of parameters is {\it invertible} on its codomain $f(\Pi)\subset \Theta$, corresponding to the set of positive couplings $ J_{ij}\geq0$,  $\forall i,j$. Thus, the two models are equivalent under such condition.

Let us start by constructing the system of equations generating the mapping for the generic case $N>2$. Following the same procedure used to obtain Eq. (\ref{dar2problem}), we find

\begin{equation}\label{darpproblem}
\makebox[\textwidth]{%
$
M_n\cdot \begin{pmatrix} 
J_{i1} \\
J_{i2} \\
J_{i3} \\
\vdots \\
J_{iN} \\
h_i \\
\end{pmatrix}\equiv
\begin{pmatrix} 
1 & 1 & ... & 1 & 1 & -1 \\
1 & 1 & ... & 1 & -1 & -1  \\
1 & 1 & ... & -1 & -1 & -1  \\
1 & ... & ... & ... & ... & -1  \\
1 & -1 & ... & -1 & -1 & -1  \\
-1 & -1 & ... & -1 & -1 & -1
\end{pmatrix}
\begin{pmatrix} 
J_{i1} \\
J_{i2} \\
J_{i3} \\
\vdots \\
J_{iN} \\
h_i \\
\end{pmatrix}
 =\frac{1}{2}
\begin{pmatrix} 
\log \left(\frac{1}{(1-\nu_i)\chi_i}-1\right) \\
\log \left(\frac{1}{\nu_i \lambda_{iN} + (1-\nu_i)\chi_i} - 1 \right)\\
\vdots \\
\vdots \\
\log \left(\frac{1}{\nu_i \sum_{j \geq 2} \lambda_{ij} + (1 - \nu_i) \chi_i} - 1 \right)\\
\log \left(\frac{1}{\nu_i+(1-\nu_i)\chi_i}-1\right) \\
\end{pmatrix}
$
}
\end{equation}

$\:\:\:\forall i=1,...,N$.

Similarly to the case $N=2$, the above system is obtained by considering $n\equiv N+1$ independent configurations for $\boldsymbol{X}_{t-1}$ and the transition probability associated with $X_t^i=1$. By matching the transition probabilities (\ref{eq:tpexp}) and (\ref{eq:tpvdar1}) associated with the $N+1$ independent configurations, one finds the system of equations (\ref{darpproblem}) for variable $i$. Then, one can repeat the same procedure for all $i$s, thus obtaining $N$ systems of $(N+1)$ linear equations in $(N+1)$ unknowns, namely $J_{i1},J_{i2},...,h_i$, each one characterized by the same matrix $M_n$ in Eq. (\ref{darpproblem}). 

Defining $\Lambda(x) \equiv \frac{e^{2x}}{1 + e^{2x}}$, the matching of probabilities associated with the $N+1$ independent configurations read as

\begin{dmath}\label{mapping}
\makebox[\textwidth]{%
$
\begin{cases}
\Lambda(h_i + \sum_{j \geq 1} J_{ij}) =  \nu_i + (1-\nu_i)\chi_i \:\: &\mbox{if} \:\: X_{t-1}^1=1, X_{t-1}^2=1, ..., X_{t-1}^N =1  \\
\Lambda(h_i - J_{i1} + \sum_{j > 1} J_{ij}) =  \nu_i (\sum_{j=2}^N \lambda_{ij}) + (1-\nu_i) \chi_i \:\: &\mbox{if} \:\: X_{t-1}^1=0, X_{t-1}^2=1, ..., X_{t-1}^N=1; \\
\Lambda(h_i - \sum_{j\leq 2}J_{ij} + \sum_{j > 2} J_{ij}) =  \nu_i (\sum_{j=3}^N \lambda_{ij}) + (1-\nu_i) \chi_i \:\: &\mbox{if} \:\: X_{t-1}^1=0, X_{t-1}^2=0, ..., X_{t-1}^N=1; \\
... & ...\\
\Lambda(h_i - \sum_{j \leq n} J_{ij}) = (1-\nu_i)\chi_i  \:\: & \mbox{if} \:\: X^1_{t-1}=0, X_{t-1}^2=0, ..., X_{t-1}^N=0,
\end{cases}$}
\end{dmath}

then, Eq. (\ref{darpproblem}) is obtained by applying $\Lambda^{-1}(y)\equiv\frac{1}{2}\log\left(\frac{1}{y}-1\right)$ to both sides of Eqs. (\ref{mapping}).


Given this result, a unique mapping $f:\Pi \rightarrow \Theta$ exists 
as long as there exists the inverse of the matrix $M_n$ in Eq. (\ref{darpproblem}), \ie if the determinant of $M_n$ is non-zero. We then start by proving the following

\begin{proposition}\label{proposition1} Given the determinant of the matrix $M_{n-1}$, then the determinant of the matrix $M_{n}$ is
\begin{equation}\label{prop1}
\det(M_{n})=(-1)^n 2\det(M_{n-1})
\end{equation}
\end{proposition}

\paragraph*{\bf Proof of Proposition \ref{proposition1}.}
By means of the  minor expansion formula (by using the minors associated with the elements of the first row), the determinant of $M_n$ can be computed as

\begin{equation}\label{proof1}
\makebox[\textwidth]{%
$
\begin{split}
\det(M_n) = &(+1) 1 \begin{vmatrix} 
1 & ... & 1 & -1 & -1  \\
1 & ... & -1 & -1 & -1  \\
 ... & ... & ... & ... & ...  \\
 -1 & ... & -1 & -1 & -1  \\
 -1 & ... & -1 & -1 & -1
\end{vmatrix} + (-1) 1 \begin{vmatrix} 
1 &  ... & 1 & -1 & -1  \\
1 & ... & -1 & -1 & -1  \\
... &  ... & ... & ... & ...  \\
1 &  ... & -1 & -1 & -1  \\
-1 &  ... & -1 & -1 & -1
\end{vmatrix}+\\
+ ... + &(-1)^n (1) \begin{vmatrix} 
1 & 1 & ... & 1 & -1  \\
1 & 1 & ... & -1 &  -1  \\
... & ... & ... & ... &  ...  \\
1 & -1 & ... & -1 &  -1  \\
-1 & -1 & ... & -1 &  -1
\end{vmatrix}+ (-1)^{n+1} (-1) \begin{vmatrix} 
1 & 1 & ... & 1 & -1   \\
1 & 1 & ... & -1 & -1   \\
... & ... & ... & ... & ... \\
1 & -1 & ... & -1 & -1  \\
-1 & -1 & ... & -1 & -1 
\end{vmatrix}.
\end{split}$}
\end{equation}

In the previous formula, one can notice that the first $n-2$ minors of the sum in the right hand side are zero, because the last two columns of each $(n-1)\times (n-1)$ matrix are indeed equal (two $(n-1)\times 1$ vectors of $-1$). Thus, Eq. (\ref{proof1}) is simplified as

\begin{equation}\label{proof2}
\det(M_n) = (-1)^n 2 \begin{vmatrix} 
1 & 1 & ... & 1 & -1   \\
1 & 1 & ... & -1 & -1   \\
... & ... & ... & ... & ...   \\
1 & -1 & ... & -1 & -1  \\
-1 & -1 & ... & -1 & -1 
\end{vmatrix} = (-1)^n 2 \det(M_{n-1})
\end{equation}

where we notice that the last two minors of (\ref{proof1}) are equal to each other and correspond to the determinant of $M_{n-1}$. Eq. (\ref{proof2}) then completes the proof of the proposition.

Thanks to this result, we are now able to prove the existence of the mapping from the VDAR(1) model to the KIM model, expressed by

\begin{proposition}\label{proposition2}
Given $\bm{\pi}\in\Pi$, there exists a solution of the problem of Eq. (\ref{darpproblem}) for any $N >0$ and this solution is unique.
\end{proposition}

\paragraph*{\bf Proof of Proposition \ref{proposition2}.}
For $N=1$, the solution can be explicitly computed as showed in Eqs. (\ref{thetaeq}) and (\ref{mueq}). For $N=2$ the problem in Eq. (\ref{darpproblem}) is equivalent to Eq. (\ref{dar2problem}) and $\det(M_3)=4$, thus there exists the inverse of the matrix $M_3$ and the solution is uniquely determined by solving the linear system of Eq. (\ref{dar2problem}). Because of Proposition \ref{proposition1}, the determinant of $M_n$ is different from zero, in particular 

\begin{dmath*}
\det(M_n)=(-1)^{\sum_{l=4}^n l}(2^{n-3})\det(M_3)=(-1)^{\sum_{l=4}^n l}(2^{n-3})4,
\end{dmath*}

$\forall n>3$ (or, equivalently, $\forall N>2$), thus resulting in the existence of the inverse matrix of $M_n$. Hence, the solution of the problem in Eq. (\ref{darpproblem}) can be uniquely determined. This completes the proof of the proposition.

\paragraph*{\bf Proof of Theorem \ref{th:equiv}.}
Given Propositions \ref{proposition1} and \ref{proposition2}, we have proved there exists a unique mapping $f:\Pi \rightarrow \Theta$ from the VDAR(1) model to the KIM. To complete the proof, we are now left with the existence of the {\it inverse} of the map $f$ in its codomain  $f(\Pi)\subset\Theta$. By restricting to such subset of parameters, the two models are equivalent.  In particular, we now prove 
the last claim of Theorem \ref{th:equiv} which states that the two models are equivalent if and only if $J_{ij} \geq 0$ $\forall i,j$. 

Let us start by proving that, if $\boldsymbol{\pi} \in \Pi$, then $J_{ij} \geq 0$. To this end let us go back to Eq. (\ref{mapping}) and notice that, combining the equations by taking the difference between the first and the second, between the second and the third and so on, we obtain the following $N$ relations

\begin{equation}\label{eq:theoproof}
\makebox[\textwidth]{%
$
\begin{cases}
\nu_i (1 - \sum_{j \geq 2} \lambda_{ij}) = \Lambda(h_i + \sum_{j \geq 2} J_{ij} + J_{i1}) - \Lambda(h_i + \sum_{j \geq 2} J_{ij} - J_{i1}) \\
 \; \; \; \; \dots \\
\nu_i \lambda_{ik} = \Lambda(h_i - \sum_{j < k} J_{ij} + \sum_{j \geq k+1} J_{ij} + J_{ik}) - \Lambda(h_i - \sum_{j < k} J_{ij} + \sum_{j \geq k+1} J_{ij} - J_{ik}) \\
 \; \; \; \;  \dots \\
\nu_i \lambda_{iN} = \Lambda(h_i - \sum_{j < N} J_{ij} + J_{iN}) - \Lambda(h_i - \sum_{j < N} J_{ij} - J_{iN}).
\end{cases}$}
\end{equation}

By definition $\nu_i \lambda_{ij} \geq 0$ for any $i,j$ (because it represents a probability), then it is
\begin{equation*}
\Lambda(C + J_{ij}) - \Lambda(C - J_{ij}) \geq 0.
\end{equation*}
Since $\Lambda(x)$ is a monotonically increasing function of $x$, the previous inequality is fulfilled if and only if $J_{ij} \geq 0$ $\forall i,j$. Thus, this condition is necessarily true if $\boldsymbol{\pi}$ is in the domain of $f$.

By following the same steps in the opposite direction it is straightforward to prove the reverse relation, that is $J_{ij} \geq 0$ is a sufficient condition to have $f^{-1}(\boldsymbol{\theta}) \in \Pi$. Indeed for any $J_{ij} \geq 0$, the product $\nu_i \lambda_{ij}$ is $0 \leq \nu_i \lambda_{ij} \leq 1$ $\forall i,j$ given the system (\ref{eq:theoproof}) and $\Lambda(x) \in [0,1]$ for any $x$. Then, by summing all the equations in system (\ref{eq:theoproof}), one obtains

\begin{equation*}
\nu_i = \Lambda(h_i + \sum_{j}J_{ij}) - \Lambda(h_i - \sum_{j}J_{ij})
\end{equation*}

which is also positive and smaller than $1$ if $J_{ij} \geq 0$ $\forall j$. Then, it follows that all the $\lambda_{ij}$ are $0 \leq \lambda_{ij} \leq 1$ $\forall i,j$. Finally, combining the first and last lines of Eq. (\ref{mapping}), one finds that $0 \leq \chi_i \leq 1$. This procedure can be repeated for all variables $i=1,...,N$, thus obtaining the  inverse mapping $f^{-1}:\Theta\vert_{J_{ij}\geq0}\rightarrow \Pi$ in the subset of the codomain $\Theta$ defined by the condition $J_{ij}\geq0$ $\forall i,j=1,...,N$. This concludes the proof.

\section{Practical implications in model inference}

\begin{figure}[t]
    \centering
    \includegraphics[width=\linewidth]{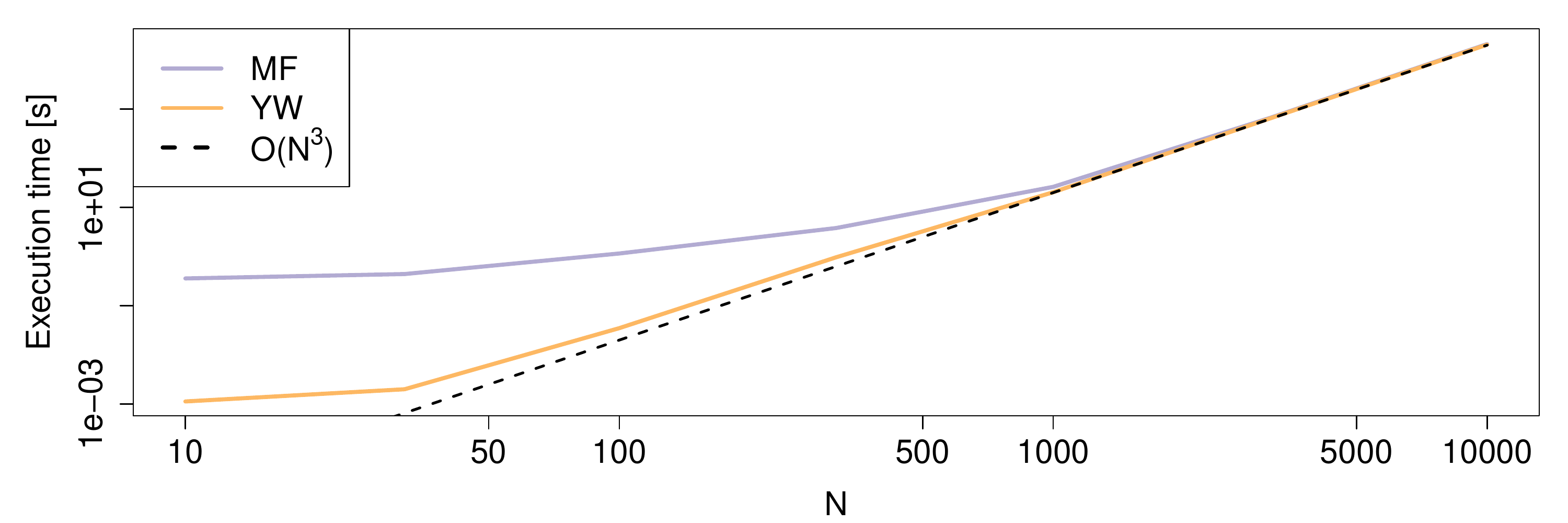}\\
    \includegraphics[width=.49\linewidth]{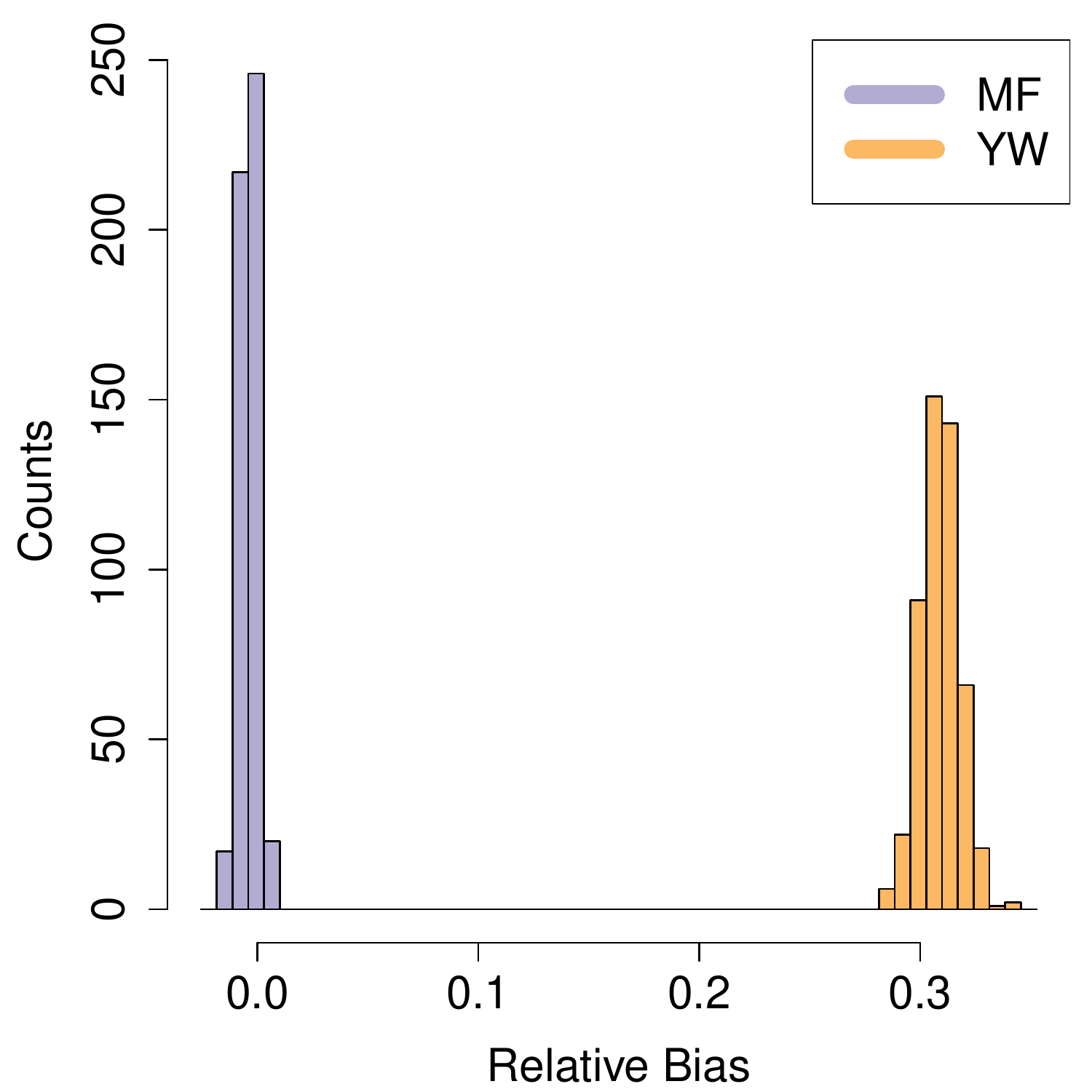}
    \includegraphics[width=.49\linewidth]{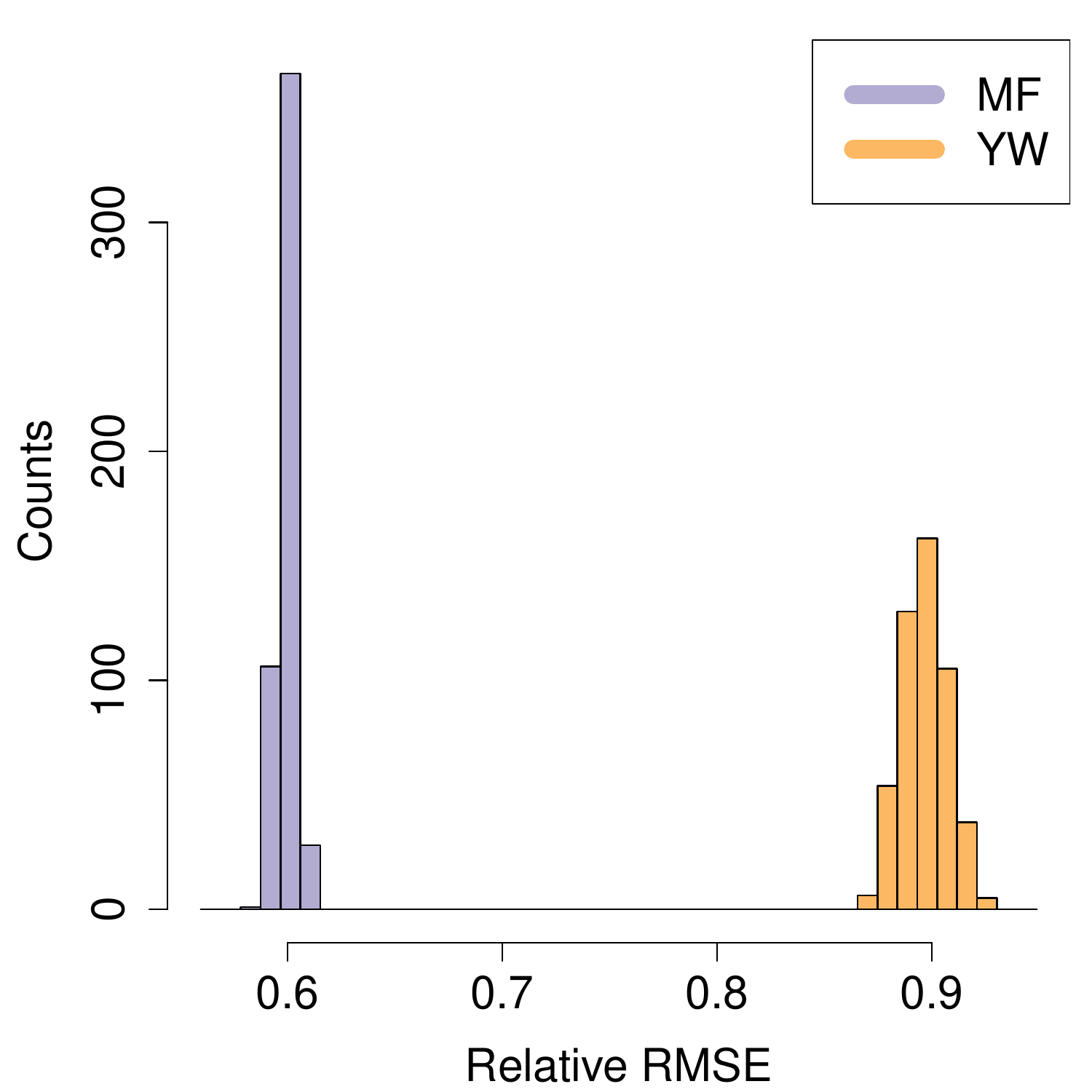}
    \caption{Comparison between the Mean Field (MF) and Yule-Walker (YW) methods for the inference of the KIM/VDAR(1). (top) Execution time varying the number of variables $N$; (bottom left) Histogram of the estimator bias relative to the average size of $J_{ij}$ over 500 simulations; (bottom right) Histogram of the estimator RMSE relative to the average size of $J_{ij}$ over 500 simulations.}
    \label{fig:YW_MF_compare}
\end{figure}

Having formally demonstrated the equivalence between the KIM and the VDAR opens the door to cross-contamination between the literatures in which they were developed. In this section we
show that one can use inference methods of the KIM developed in the statistical literature, namely the mean field method \cite{mezard2011exact},  for improving standard inference methods for discrete autoregressive processes, namely the Yule-Walker equations.

Specifically, a popular method in time series literature for the inference of autoregressive models is the method of moments, which generates the so-called Yule-Walker equations matching the empirically measured moments with the ones implied by the model parameters \cite{tsay2014financial}. In the context of the VDAR(1) it can be shown \cite{mazzarisi2020tail} that the Yule-Walker equations read

\begin{eqnarray}
    \mathbb{E}(\mathbf{X}_t) = \boldsymbol{\phi} + \boldsymbol{\Psi} \mathbb{E}(\mathbf{X}_{t-1}) \\
    \mathbb{E}(\mathbf{\tilde{X}_t \tilde{\mathbf{X}}_{t-1}}) = \boldsymbol \Psi \mathbb{E}(\tilde{\mathbf{X}}_{t-1} \tilde{\mathbf{X}}_{t-1})
\end{eqnarray}

where $\tilde{\mathbf{X}}_t = \mathbf{X}_t - \mathbb{E}(\mathbf{X}_t)$, $\boldsymbol \phi$ is a $N$-dimensional vector and $\boldsymbol\Psi$ is a $N \times N$ matrix. Solving these linear systems for $\boldsymbol\phi$ and $\boldsymbol\Psi$ allows to obtain the VDAR parameters as $\nu_i = \sum_j \boldsymbol\Psi_{ij}$, $\lambda_{ij} = \boldsymbol\Psi_{ij}/\nu_i$ and $\chi_i = \boldsymbol\phi_i/(1 - \nu_i)$.

On the other hand, the statistical mechanics literature has developed suitable approximation methods  of maximum likelihood estimation of KIM. In particular, we consider the method developed by M\'{e}zard and Sakellariou \cite{mezard2011exact}, which takes advantage of a mean field approximation to infer the parameters of the Kinetic Ising Model. The method is exact if the $J$ generating the data has all $J_{ij} \neq 0$ and can be adapted to a sparse version through $\ell_1$-regularization or decimation methods \cite{decelle2015inference}. It has been extensively used in applications of the model to real financial and neural data \cite{campajola2020modelling, roudi2015multi}.

The equivalence between KIM and VDAR we proved in this paper allows us to use the mean field method for doing inference on a VDAR model. To test this idea, we apply mean field and Yule-Walker method on simulated data and compare them both in speed and accuracy. The speed is measured by the time needed to perform the inference as a function of $N$ on a regular commercial laptop, while the accuracy is measured by the bias and root mean squared error (RMSE) of the estimator of $J$ over 500 simulations. We simulate the KIM with uniformly distributed $J_{ij} \sim \mathcal{U}(1/2N, 1/N)$ to keep the model far from the dynamic ferromagnetic transition \cite{crisanti1988dynamics}, for $T = 10N$ time steps and $N$ ranging from $10$ to $10,000$. For the sake of simplicity we consider $h_i = 0 \, \forall \, i$ in our simulations. We show the results of the comparison in Figure \ref{fig:YW_MF_compare}, where it is clear that the Yule-Walker method is faster on small scale systems but is also less accurate. In particular we see that the method of moments has a positive bias and a relatively large RMSE, whereas the mean field method has close to zero bias and a smaller RMSE. The computational effort required in the two methods is comparable as the biggest contribution comes from the inversion of the covariance matrix, typically achieved in $O(N^3)$ operations, hence they present similar execution time for large $N$. Thus the mean field method is a better choice for the inference of the KIM/VDAR(1), and this result is somewhat expected, since mean field is a likelihood-based method whereas the Yule-Walker equations are not. 

This simple analysis shows that the equivalence between the two models can be leveraged to identify inference methods, originally developed for the KIM, that can be used in the inference of the VDAR. As shown above, this can lead to significant  improvements in performance when compared to standard VDAR inference methods.    

\section{Conclusions}\label{sec:conclusion}
In conclusion, the VDAR(1) model is equivalent to the Kinetic Ising Model thanks to the existence of a unique mapping for both the binary random variables and the parameters as long as the $\bm{J}$ parameters of the Kinetic Ising Model are positive or zero, as a consequence of the fact that the $\bm{\nu}$ and $\bm{\lambda}$ parameters only account for non-negative lagged correlations among random variables. Moreover, since the two models can be interpreted as the maximum entropy distribution of binary random sequences with given  means, and both auto- and  cross-correlations (only non-negative correlations for the specific case of the VDAR(1) model), both of them represent further the best choice in describing such binary random sequences in absence of prior information on other metrics, according to the principle of maximum entropy.

There are several directions in which future research can go to take advantage of our equivalence theorem. We have already shown that inference methods developed in the statistical physics literature can be used to improve model estimation; another straightforward application of this theorem is that of defining an extension of the VDAR(1) including negative correlations, as the equivalent to the KIM without the restriction on the positive $J$ elements. Finally, it is common in autoregressive models to consider Markov chains of order higher than 1, that is where there are explicit parameters linking the value of $X_t$ to the value of $X_{t-k}$ with $k = 1, \dots, p$, $p > 1$. The properties of a Kinetic Ising Model with higher order interactions have not been studied yet to the best of our knowledge, thus opening another interesting perspective to explore in the context of this cross-contamination between very active research fields.

\medskip

\section*{Acknowledgments} FL acknowledges partial support by the European Integrated Infrastructure for Social Mining and Big Data Analytics (SoBigData++, Grant Agreement \#871042). DT acknowledges GNFM-Indam for financial support.

\bibliographystyle{unsrt}
\bibliography{biblio.bib}

\begin{thebibliography}{10}

\bibitem{calcagnile2018collective}
Lucio~Maria Calcagnile, Giacomo Bormetti, Michele Treccani, Stefano Marmi, and
  Fabrizio Lillo.
\newblock Collective synchronization and high frequency systemic instabilities
  in financial markets.
\newblock {\em Quantitative Finance}, 18(2):237--247, 2018.

\bibitem{hong2009granger}
Yongmiao Hong, Yanhui Liu, and Shouyang Wang.
\newblock Granger causality in risk and detection of extreme risk spillover
  between financial markets.
\newblock {\em Journal of Econometrics}, 150(2):271--287, 2009.

\bibitem{bouchaud2002statistical}
Jean-Philippe Bouchaud, Marc M{\'e}zard, Marc Potters, et~al.
\newblock Statistical properties of stock order books: empirical results and
  models.
\newblock {\em Quantitative finance}, 2(4):251--256, 2002.

\bibitem{fredrickson1984kinetic}
Glenn~H Fredrickson and Hans~C Andersen.
\newblock Kinetic {Ising} model of the glass transition.
\newblock {\em Physical review letters}, 53(13):1244, 1984.

\bibitem{jacobs1978discrete}
Patricia~A Jacobs and Peter~AW Lewis.
\newblock Discrete time series generated by mixtures. {III}. autoregressive
  processes ({DAR} (p)).
\newblock Technical report, Naval Postgraduate School Monterey Calif, 1978.

\bibitem{jacobs1983stationary}
Patricia~A Jacobs and Peter~AW Lewis.
\newblock Stationary discrete autoregressive-moving average time series
  generated by mixtures.
\newblock {\em Journal of Time Series Analysis}, 4(1):19--36, 1983.

\bibitem{mazzarisi2020tail}
Piero Mazzarisi, Silvia Zaoli, Carlo Campajola, and Fabrizio Lillo.
\newblock Tail granger causalities and where to find them: Extreme risk
  spillovers vs spurious linkages.
\newblock {\em Journal of Economic Dynamics and Control}, 121:104022, 2020.

\bibitem{schneidman2006weak}
Elad Schneidman, Michael~J Berry, Ronen Segev, and William Bialek.
\newblock Weak pairwise correlations imply strongly correlated network states
  in a neural population.
\newblock {\em Nature}, 440(7087):1007--1012, 2006.

\bibitem{marre2009prediction}
Olivier Marre, Sami El~Boustani, Yves Fr{\'e}gnac, and Alain Destexhe.
\newblock Prediction of spatiotemporal patterns of neural activity from
  pairwise correlations.
\newblock {\em Physical review letters}, 102(13):138101, 2009.

\bibitem{jaynes1982rationale}
Edwin~T Jaynes.
\newblock On the rationale of maximum-entropy methods.
\newblock {\em Proceedings of the IEEE}, 70(9):939--952, 1982.

\bibitem{presse2013principles}
Steve Press{\'e}, Kingshuk Ghosh, Julian Lee, and Ken~A Dill.
\newblock Principles of maximum entropy and maximum caliber in statistical
  physics.
\newblock {\em Reviews of Modern Physics}, 85(3):1115, 2013.

\bibitem{marcaccioli2020correspondence}
Riccardo Marcaccioli and Giacomo Livan.
\newblock Correspondence between temporal correlations in time series, inverse
  problems, and the spherical model.
\newblock {\em Phys. Rev. E}, 102:012112, Jul 2020.

\bibitem{sastre2003nominal}
Francisco Sastre, Ivan Dornic, and Hugues Chat{\'e}.
\newblock Nominal thermodynamic temperature in nonequilibrium kinetic {Ising}
  models.
\newblock {\em Physical review letters}, 91(26):267205, 2003.

\bibitem{derrida1987exactly}
Bernard Derrida, Elizabeth Gardner, and Anne Zippelius.
\newblock An exactly solvable asymmetric neural network model.
\newblock {\em EPL (Europhysics Letters)}, 4(2):167, 1987.

\bibitem{crisanti1988dynamics}
A~Crisanti and Haim Sompolinsky.
\newblock Dynamics of spin systems with randomly asymmetric bonds: {Ising}
  spins and {Glauber} dynamics.
\newblock {\em Physical Review A}, 37(12):4865, 1988.

\bibitem{sides1998kinetic}
SW~Sides, PA~Rikvold, and MA~Novotny.
\newblock Kinetic {Ising} model in an oscillating field: Finite-size scaling at
  the dynamic phase transition.
\newblock {\em Physical review letters}, 81(4):834, 1998.

\bibitem{ising1925beitrag}
Ernst Ising.
\newblock Beitrag zur theorie des ferromagnetismus.
\newblock {\em Zeitschrift f{\"u}r Physik A Hadrons and Nuclei},
  31(1):253--258, 1925.

\bibitem{edwards1975theory}
Samuel~Frederick Edwards and Phil~W Anderson.
\newblock Theory of spin glasses.
\newblock {\em Journal of Physics F: Metal Physics}, 5(5):965, 1975.

\bibitem{kirkpatrick1978infinite}
Scott Kirkpatrick and David Sherrington.
\newblock Infinite-ranged models of spin-glasses.
\newblock {\em Physical Review B}, 17(11):4384, 1978.

\bibitem{roudi2011mean}
Yasser Roudi and John Hertz.
\newblock Mean field theory for nonequilibrium network reconstruction.
\newblock {\em Physical review letters}, 106(4):048702, 2011.

\bibitem{capone2015inferring}
Cristiano Capone, Carla Filosa, Guido Gigante, Federico Ricci-Tersenghi, and
  Paolo Del~Giudice.
\newblock Inferring synaptic structure in presence of neural interaction time
  scales.
\newblock {\em PloS one}, 10(3):e0118412, 2015.

\bibitem{imparato2007ising}
A~Imparato, A~Pelizzola, and M~Zamparo.
\newblock Ising-like model for protein mechanical unfolding.
\newblock {\em Physical review letters}, 98(14):148102, 2007.

\bibitem{coolen1993dynamics}
ACC Coolen and D~Sherrington.
\newblock Dynamics of fully connected attractor neural networks near
  saturation.
\newblock {\em Physical review letters}, 71(23):3886, 1993.

\bibitem{dunn2013learning}
Benjamin Dunn and Yasser Roudi.
\newblock Learning and inference in a nonequilibrium {Ising} model with hidden
  nodes.
\newblock {\em Physical Review E}, 87(2):022127, 2013.

\bibitem{campajola2019inference}
Carlo Campajola, Fabrizio Lillo, and Daniele Tantari.
\newblock Inference of the kinetic {Ising} model with heterogeneous missing
  data.
\newblock {\em Physical Review E}, 99(6):062138, 2019.

\bibitem{bouchaud2013}
Jean-Philippe Bouchaud.
\newblock Crises and collective socio-economic phenomena: Simple models and
  challenges.
\newblock {\em Journal of Statistical Physics}, 151(3):567--606, May 2013.

\bibitem{campajola2020unveiling}
Carlo Campajola, Fabrizio Lillo, and Daniele Tantari.
\newblock Unveiling the relation between herding and liquidity with trader
  lead-lag networks.
\newblock {\em Quantitative Finance}, 20(11):1765--1778, 2020.

\bibitem{campajola2020modelling}
Carlo Campajola, Domenico Di~Gangi, Fabrizio Lillo, and Daniele Tantari.
\newblock Modelling time-varying interactions in complex systems: the {Score
  Driven Kinetic Ising Model}.
\newblock {\em arXiv preprint arXiv:2007.15545}, 2020.

\bibitem{jaynes1957information}
Edwin~T Jaynes.
\newblock Information theory and statistical mechanics.
\newblock {\em Physical review}, 106(4):620, 1957.

\bibitem{barnett2013information}
Lionel Barnett, Joseph~T Lizier, Michael Harr{\'e}, Anil~K Seth, and Terry
  Bossomaier.
\newblock Information flow in a kinetic {Ising} model peaks in the disordered
  phase.
\newblock {\em Physical review letters}, 111(17):177203, 2013.

\bibitem{hanneke2010discrete}
Steve Hanneke, Wenjie Fu, Eric~P Xing, et~al.
\newblock Discrete temporal models of social networks.
\newblock {\em Electronic Journal of Statistics}, 4:585--605, 2010.

\bibitem{park2004statistical}
Juyong Park and Mark~EJ Newman.
\newblock Statistical mechanics of networks.
\newblock {\em Physical Review E}, 70(6):066117, 2004.

\bibitem{dehnert2003discrete}
Manuel Dehnert, WE~Helm, and M-Th H{\"u}tt.
\newblock A discrete autoregressive process as a model for short-range
  correlations in {DNA} sequences.
\newblock {\em Physica A: Statistical Mechanics and its Applications},
  327(3-4):535--553, 2003.

\bibitem{kim2008mean}
Jeongsim Kim, Bara Kim, and Khosrow Sohraby.
\newblock Mean queue size in a queue with discrete autoregressive arrivals of
  order p.
\newblock {\em Annals of Operations Research}, 162(1):69--83, 2008.

\bibitem{williams2019effects}
Oliver~E Williams, Fabrizio Lillo, and Vito Latora.
\newblock Effects of memory on spreading processes in non-markovian temporal
  networks.
\newblock {\em New Journal of Physics}, 21(4):043028, 2019.

\bibitem{taranto2014adaptive}
D~E Taranto, Giacomo Bormetti, and Fabrizio Lillo.
\newblock The adaptive nature of liquidity taking in limit order books.
\newblock {\em Journal of Statistical Mechanics: Theory and Experiment},
  2014(6):P06002, 2014.

\bibitem{mazzarisi2020dynamic}
Piero Mazzarisi, Paolo Barucca, Fabrizio Lillo, and Daniele Tantari.
\newblock A dynamic network model with persistent links and node-specific
  latent variables, with an application to the interbank market.
\newblock {\em European Journal of Operational Research}, 281(1):50--65, 2020.

\bibitem{mezard2011exact}
Marc M{\'e}zard and J~Sakellariou.
\newblock Exact mean-field inference in asymmetric kinetic ising systems.
\newblock {\em Journal of Statistical Mechanics: Theory and Experiment},
  2011(07):L07001, 2011.

\bibitem{tsay2014financial}
Ruey~S Tsay.
\newblock Financial time series.
\newblock {\em Wiley StatsRef: Statistics Reference Online}, pages 1--23, 2014.

\bibitem{decelle2015inference}
Aur{\'e}lien Decelle and Pan Zhang.
\newblock Inference of the sparse kinetic {Ising} model using the decimation
  method.
\newblock {\em Physical Review E}, 91(5):052136, 2015.

\bibitem{roudi2015multi}
Yasser Roudi, Benjamin Dunn, and John Hertz.
\newblock Multi-neuronal activity and functional connectivity in cell
  assemblies.
\newblock {\em Current opinion in neurobiology}, 32:38--44, 2015.

\end{thebibliography}

\end{document}